\documentclass[12pt]{article}

\usepackage[margin=0.5in]{geometry}
\usepackage{amsfonts}
\usepackage{amsmath}
\usepackage{lscape}
\usepackage{graphicx}
\usepackage{epstopdf}
\usepackage{color}
\usepackage{float} % Enables [H] to fix location of floats
\usepackage{arydshln} % Provides verticle dashed line in tables
\usepackage[lined,boxed,commentsnumbered]{algorithm2e} % Enables algorithm table
\usepackage{caption} % Enables 2X2 figures
\usepackage{subcaption} % Enables 2X2 figures
\usepackage{amsthm} %Enables theorem environment
\usepackage{multirow} %Enables merging cells across rows
\usepackage[round]{natbib} %Enables simple routine citiations which are common to JASA

\usepackage{secdot}
\usepackage{amssymb}
\usepackage{setspace} %I am using this to single space the algorithm
\usepackage[normalem]{ulem} % This enables strikeout (\sout)
\newtheorem{remark}{Remark}

\begin{document}

\title{Penalized Component Hub Models}
\date{}
\author{
    Charles Weko\\
    Headquarters, Department of the Army\\
    United States Army\\
    charles.w.weko.mil@mail.mil 
		  \and
    Yunpeng Zhao\\
    Department of Statistics\\
    George Mason University\\
    Yunpeng.Zhao@asu.edu 
}

\maketitle

\newpage
\begin{abstract}
	
Social network analysis presupposes that observed social behavior is influenced by an unobserved network.  Traditional approaches to inferring the latent network use pairwise descriptive statistics that rely on a variety of measures of co-occurrence.  While these techniques have proven useful in a wide range of applications, the literature does not describe the generating mechanism of the observed data from the network.

In a previous article, the authors presented a technique which used a finite mixture model as the connection between the unobserved network and the observed social behavior. This model assumed that each group was the result of a star graph on a subset of the population.  Thus,  each group was the result of a leader who selected members of the population to be in the group.  They called these \textit{hub models}.  

This approach treats the network values as parameters of a model. However, this leads to a general challenge in estimating parameters which must be addressed. For small datasets there can be far more parameters to estimate than there are observations.  Under these conditions, the estimated network can be unstable.

In this article, we propose a solution which penalizes the number of nodes which can exert a leadership role.  We implement this as a \textit{pseudo-Expectation Maximization} algorithm.

We demonstrate this technique through a series of simulations which show that when the number of leaders is sparse, parameter estimation is improved.  Further, we apply this technique to a dataset of animal behavior and an example of recommender systems.

\end{abstract}


\section{Introduction}

Networks consist of discrete \textit{nodes} or \textit{verticies} which are connected by \textit{links} or \textit{edges}. These pairwise connections are frequently represented by a square matrix called an \textit{adjacency matrix}.  Network analysis has drawn attention in a wide variety of scientific and engineering disciplines because of the practicality of the network structure. The applications of networks include concrete problems such as finding the shortest path through a transportation system or determining the maximum flow through an electrical transmission system \citep{Hiller01}.  The generality of networks allows for their application to more abstract problems such as the propagation of disease or information through a population \citep{Jackson08}.  Applications further extend to identifying key nodes in social networks \citep{Koschützki2005}, community detection among weblogs on the World Wide Web \citep{Karrer10}, link prediction in social and biological networks \citep{Liben2007,Zhao2013}, as well as many others \citep{Kolaczyk09,Goldenberg10,Newman11}.

Traditionally, statistical network analysis focuses on modeling the random generation of observed or \textit{explicit} network structure. For physical networks, like communication systems or railway networks, the nodes are clearly defined and the links between nodes can be directly observed \citep{Hiller01, Kolaczyk09, Newman11}. 

In other fields of research, explicit network structure may not be observable.  This is especially true in the social sciences where the observed raw data is usually the social behavior instead of an explicit network structure \citep{Freeman89,Whitehead08}. This situation may also occur in the analysis of protein-protein interaction or gene regulatory networks. In these situations, the observed behavior is presumed to result from a latent network structure. For instance, researchers may not directly observe ``friendships'' within a population; instead, they may observe some social behavior (e.g., four people gather together with a certain frequency or they visited each other's house at least once in a month). 

The notion that there is a connection between observable behavior and network structure can be traced to the so-called \textit{social network perspective} proposed by \citet{Moreno34}. \citet{Wasserman94} also gave a detailed explanation of this concept. The central principle of the social network perspective is that a network model governs the action of individual nodes and makes them behave interdependently.  This relationship between behavior and network structure suggests that the network may be inferred from such observed behavior. In a previous article, \citet{Zhao16} developed a model which used the network as a parameter for the random generation of observed behavior. 

% page 4 of Wasserman

The construction of latent networks often relies on data structures generated from surveys in which individuals or researchers report  relationships \citep{Sampson69,Zachary77}.  In this article we focus on an alternative type of dataset which is frequently collected in the social sciences and which can be generalized to other areas of research. \citet{Wasserman94} introduce such a dataset using the example of children attending birthday parties. In Table \ref{T:Birthday}, the value 1 indicates that a specific child attended a party, and 0 indicates otherwise. For example, Allison attended Parties 1 and 3 but did not attend Party 2.  \citet{Whitehead08} refers to each party as a \textit{group} and Table \ref{T:Birthday} as a \textit{group-by-individual matrix}.  %\citet{Zhao15} 
\citet{Zhao16} referred to this type of data as \textit{grouped data}. 

\begin{table}
\caption{\small Dataset for six children and three birthday parties, Adapted from \citep{Wasserman94}.}
\label{T:Birthday}
\begin{center}
\begin{tabular}{|c  | c c c c c c |}
\hline
  &\multicolumn{6}{c|}{Child} \\
Party & Allison & Drew & Eliot & Keith & Ross & Sarah \\
\hline
1 & 1 & 0 & 0 & 0 & 1 & 1\\
2 & 0 & 1 & 1 & 0 & 1 & 1\\
3 & 1 & 0 & 1 & 1 & 1 & 0\\
\hline
\end{tabular}
\end{center}
\end{table}

The existing methods for network inference from grouped data are essentially descriptive statistics. The most common approach is to use the frequency of co-occurrence between two nodes to estimate the strength of the link between individuals \citep{Zachary77,Freeman89,Wasserman94,Kolaczyk09}. We refer to this measure as the \textit{co-occurrence matrix}. As an alternative, the \textit{half weight index} \citep{Dice45,Cairns87,Bejder98,Whitehead08} estimates the strength of the link by the frequency that two nodes co-occur given that one of them is observed. 

One shortcoming of these techniques is that they do not define how the observed data is generated from the estimated statistics. A particular challenge is that the probability of co-occurrence is not equivalent to the probability of connection.  For example, in Table \ref{T:Birthday} it is possible that two children who do not know each other attended the same party because they are invited by a mutual friend.  It remains unclear what model assumption justifies the network structure inferred by these measures.

%\citet{Zhao15} 
\citet{Zhao16} proposed a simplistic generating mechanism for grouped data based on a network structure. The \textit{hub model} (HM) assumes that each observed group is the result of a leader bringing together a subset of the population.  That is, every group is brought together by a central node (often referred to as the \textit{leader}).  The other members of the group are present based on their relationship to this leader. Thus, the hub model parameters have an interpretation which can be easily applied to relevant research questions. 

Despite the fact that hub models assume an intuitive generating mechanism and perform well with sufficient observations, the number of parameters in the model presents a challenge. If we let $n$ be the number of nodes in the network, the network contains $O(n^2)$ parameters. Therefore, a moderate-sized network (e.g., $n=50$) would in principle require a large number of observations to accurately estimate the network. 

Moreover, in most practical situations, the central node of each group is unobserved. Without any prior information, it is possible for any node in the group to be the central node for that group.  %\citet{Zhao15}
\citet{Zhao16} use an Expectation-Maximization (EM) algorithm to identify the central node for each group.  As $n$ increases, the possibility for larger and larger groups also increases.  Thus identifying the central node of such groups can be difficult because there are many nodes which could be central and the probability of each node being central can be small.

In practice, it is not necessary to model every node in the population as a potential leader. For example, there may be low ranking members of the population who do not have the authority or influence to initiate a group.  This is especially true when the number of observations is small. Therefore, we propose a \textit{penalized component hub model} (PCHM) to reduce the hub model's complexity. Using a penalized likelihood of hub models, the probability that a node is a leader is shrunk towards 0 when that probability is small. 

The PCHM assumes \textit{sparse} parameters.  That is, only a small proportion of the nodes have a non-zero probability of forming a group. Since the hub model is an example of a finite mixture model, we essentially penalize the number of components in the mixture model. 

This penalization technique belongs to the class of regularization methods which have been extensively studied in the statistical literature. For example, least absolute shrinkage and selection operator (LASSO) introduced by \citet{Tibshirani96} is a famous $L_1$ regularization method for variable selection in linear regression. Ridge regression \citep{Hoerl1} applies $L_2$ regularization to reduce the variance of the coefficients estimates and hence obtains smaller mean square error than least square estimates. Similarly in the PCHM case, regularization on the probabilities of nodes being centers leaders increases the stability of the estimated networks and yields better performance when the sample size is limited. 

Regularization techniques have been widely used in graphical models and covariance estimation to obtain a ``sparse'' estimated adjacency matrix \citep{Bickel08,Friedman08,Guo10}. However, the definition of ``sparse'' in these techniques is different from the definition we will use.  Traditional techniques define the network structure solely based on an adjacency matrix.  Thus a ``sparse'' network is one where the adjacency matrix contains many elements which are equal to zero.  In this case, regularization of the network is achieved by penalizing the elements of the adjacency matrix of the network.  

Hub models define the network structure using two parameters (a mixing distribution and an adjacency matrix).  Under PCHM, sparsity is defined on the mixing distribution.  Thus PCHM penalizes the probability of nodes being centers. The detailed explanation motivating this approach will be given in Section \ref{S:Review} and further elaborated in Section \ref{S:Methodology}.

The rest of this article is organized as follows. We start with a brief review of hub models to motivate our approach in Section \ref{S:Review}. We propose PCHM and the algorithm for solving the penalized likelihood in Section \ref{S:Methodology}. In Section \ref{S:Selecting Tuning Parameters}, we discuss the application of the Bayesian Information Criterion (BIC) for tuning parameter selection. Simulation studies are provided in Section \ref{S:Simulation}. In Section \ref{S:Data}, we apply the PCHM to a dataset of Hector's dolphins \citep{Bejder98} and a recommender system in supplemental materials.

\section{Motivation for Penalizing Hub Models} \label{S:Review}

The technique of penalizing hub models is closely related to the technique used to solve hub models.  Therefore, we introduce the motivation concurrently with the basic notation of hub models.

\subsection{Data}

For a population of $n$ individual nodes, $V=\{v_1,\dots, v_n\}$, we observe  $T$ subsets of the global population, $\{V^{(t)}|V^{(t)} \subseteq V, t=1,...,T \}$.  
Each observed subset $V^{(t)}$ can be represented by an $n$ length row vector $G^{(t)}$ where: for $i=1,...,n$, 

\[ G_i^{(t)} = \left\{ 
   \begin{array}{l l}
     1 & \quad \textnormal{if $v_i\in V^{(t)}$},\\
     0 & \quad \textnormal{if $v_i \notin V^{(t)}$}.
   \end{array} \right.\]

The full set of observations is denoted by a $T \times n$ matrix, $G$.  The $t^{th}$ row of $G$ is $G^{(t)}$ . 

\subsection{Empirical Methods} \label{S:empirical}
The classical approaches to infer the latent network in the social sciences literature rely on descriptive statistics. We give the formula of two popular techniques, \textit{co-occurrence matrix} and \textit{half weight index}, which we will compare with PCHM in data analysis in Section \ref{S:Data}.

A co-occurrence matrix, $O$, is an $n\times n $ symmetric matrix, defined as:  
\begin{equation}
	O=\frac{G' G}{T},\notag
\end{equation}
where $O_{ij}$ is the relative frequency with which nodes $v_i$ and $v_j$ are observed in the same group.

The half weight index, $H$, is an $n\times n $ symmetric matrix whose elements, $H_{ij}$, estimate the conditional probability that the nodes $v_i$ and $v_j$ are observed in the same group given that one of them is observed. It has been introduced in a number of equivalent forms \citep{Dice45,Cairns87}. We give the simplest form as follows, 
\begin{equation}\label{E:HWI1}
H_{ij} = \frac{2\sum_{t=1}^T G_i^{(t)} G_j^{(t)}}{\sum_{t=1}^T G_i^{(t)}+\sum_{t=1}^T G_j^{(t)}}.\notag
\end{equation}

\subsection{Generating Mechanism of Hub Models}\label{S: GenMech}

The empirical methods do not establish the connection of the generating mechanism of the groups and these descriptive statistics. Hub models introduce a model-based approach for network estimation. Hub models assume that at the moment of observation each group is a brought together by a single leader. 

The central node initiating $G^{(t)}$ is represented by an $n$ length row vector, $S^{(t)}$, where

\[ S_i^{(t)} = \left\{ 
   \begin{array}{l l}
     1 & \quad \text{if $v_i$ the central node of sample $t$},\\
     0 & \quad \text{otherwise}.
   \end{array} \right.\]
   
There is one and only one element of $S^{(t)}$ that is equal to 1. 

Let $S$ be a $T \times n$ matrix, where the $t^{th}$ row of $S$ is $S^{(t)}$. $S$ is only observable in a narrow range of datasets, for example, emails \citep{Michalski14} and Congressional legislation \citep{Fowler06a}.  We focus on the more general case where $S$ is unobserved.

Under the hub model, each group $G^{(t)}$ is independently generated by the following two step process. 

\begin{enumerate}
	\item The central node is drawn from a multinomial distribution with parameter $\rho=(\rho_1,...,\rho_n)$, where $\rho_i=\mathbb{P}(S_i^{(t)}=1)$, with the constraint $\sum_i \rho_i=1$.  Thus, $\rho_i$ represents the probability that node $v_i$ will form a group.

	\item The central node, $v_i$, includes $v_j$ in the group with probability $A_{ij}$, where, $A_{ij}=	\mathbb{P}(G_j^{(t)}=1|S_i^{(t)}=1)$.  
\end{enumerate}
Thus, $\{A,\rho\}$ is the set of parameters of hub models. 

It is assumed that $A_{ii}=1$ for all $i$.  This means that the central node will always include itself in any group it forms. As a result, the only way that $v_i$ can appear as a singleton is if $v_i$ is the center.

Hub models provide researchers with parameters which are easy to interpret.  The value of $\rho_i$ indicates the influence that the node $v_i$ exerts over the population.  Thus, if $\rho_x=0.2$ and $\rho_y=0.1$, we could say that node $v_x$ has twice the influence of node $v_y$.  At the same time, if $\rho_z=0.0001$, we could say that node $v_z$ is a subordinate member of the population and exerts negligible influence of its own.  

The value of $A_{ij}$ indicates the ``popularity'' of node $v_j$ with node $v_i$.  If $A_{ij}$ is close to one, this means that whenever node $v_i$ forms a group that he chooses to include node $v_j$.  However, when $A_{ij}$ is close to zero this indicates that there is antipathy between the individuals.

A finite mixture model of multivariate Bernoulli random variables like hub models is not identifiable \citep{Teicher61, Carreira00}.  %\citet{Zhao15}
\citet{Zhao16} showed that symmetry of the adjacency matrix (i.e., $A_{ij}=A_{ji}$) is a sufficient condition ensuring identifiability.  We will use the same symmetric assumption in this article. 

\subsection{Discussion of Hub Models}

Now that we have introduced the basic notation of hub models, it is worthwhile to explain why we would study hub models and what their implications would be.

We begin with why we are interested in a model-based approach for network estimation from grouped data. There is an extensive literature of statistical analysis on social networks. However, this literature usually models the behavior of given adjacency matrices. Examples of such models include stochastic blockmodels \citep{Holland83,Snijders&Nowicki1997,Nowicki2001}, latent space models \citep{Hoff2002}, preferential attachment model \citep{Barabasi&Albert1999}, and so on. Statistical methods concerning implicit network and grouped data focus on randomizing the existing data to test the sensitivity of the descriptive statistics \citep{Whitehead08}.  They do not propose stochastic processes for data generation. To the best of our knowledge, hub models are the first approach which fills this gap.

Hub models formally link the underlying network structure to the observed group data.  Traditional descriptive statistics which estimate network structure answer the question ``given a set of data, what is the underlying structure''.  However, these statistics cannot be used to answer the reverse question, ``given an underlying structure, generate a `similar' dataset''.  Hub models are an important reinforcement of the notion of social networks because they are able to answer both questions.

The assumption of hub models comes from the observation that certain types of groups such as e-mails and Congressional legislation explicitly have this structure.  In these cases, the leader of the group is usually known; however, this structure can exist in situations where the leader is unknown.  For example, office meetings are generally called by a single manager who exerts control over subordinate employees.  Also, journal articles typically have a primary author who coordinates and leads a group of collaborators. In all of these examples, the leader may not be explicit, but the structure of the groups still follows the hub model.

\subsection{Motivation of Penalized Component Hub Models}\label{S:PCHM Motivation}

The hub model contains $\big[\frac{n(n-1)}{2}+(n-1)\big]$ free parameters due to $A$ and $\rho$.  In principle this demands  a large number of observations to estimate the parameters accurately. If the sample size is moderate, it would be helpful to reduce the number of parameters. 

The hub model's complexity can be significantly reduced if only a small portion of nodes could be central nodes. That is, if $\rho_i=0$ for a non-trivial number of nodes.  Let $n_o$ denote the number of $\rho_i$'s which are non-zero. Without loss of generality, we define \textit{sparsity} as $\rho=(\rho_1, \rho_2,...,\rho_{(n_o)},0,...,0)$ and refer to the set of nodes with $\rho_i>0$ as the \textit{leader set}.

Since $\rho$ can be interpreted as the mixing distribution of a finite mixture model, $\rho_i=0$ implies that the elements of the adjacency matrix $A_{ij}$ for all $j$ have no impact on the likelihood function. Essentially the adjacency matrix has the dimension $n_o\times n$.  However, in order to be able to compare models with different $n_o$ conveniently, we want to define $A$ as $n \times n$ matrix.  Therefore, if both $\rho_i=0$ and $\rho_j=0$, then we set $A_{ij}=0$ as a convention and apply the symmetry assumption.  Thus, the structure of the adjacency matrix becomes 

\begin{center}
$\begin{array}{c|c} 
   X & Y \\
   \hline
   Y' & \textbf{0}
\end{array}$
\end{center}

where $X$ is an $n_o \times n_o$ symmetric matrix, $Y$ is an $n_o \times (n-n_o)$ matrix, and \textbf{0} is an $(n-n_o) \times (n-n_o)$ matrix of zeros. There are $\frac{n_o(n_o-1)}{2}$ elements in $X$ and $n_o(n-n_o)$ elements in $Y$ to be estimated. 

Our goal is to propose a regularization method to shrink $\rho_i$ with a small value towards 0 and obtain a sparse estimator  $\hat{\rho}$ while maintaining the constraint $\sum_i \hat{\rho}_i =1$.  

\begin{remark}

At first glance, a more direct approach to parameter reduction would be to penalize $A_{ij}$ directly using $L_1$ regularization. However, this approach is not appropriate for hub models because there is a conceptual difference between the matrices in hub models and classical graphical models. 

In any situation where variable selection is at issue, there is a ``simplest version'' of the model which indicates the case that the response variable is independent of the model parameters.

As an example, the parameter of interest in graphical models is a covariance matrix, $\Omega$, or inverse covariance matrix, $\Omega^{-1}$.  Thus, $\Omega^{-1}_{ij}=0$ if variables $i$ and $j$ are conditionally independent and the diagonal matrix is the ``simplest version'' for a graphical model. Therefore, $L_1$ regularization penalizes $\Omega^{-1}$ towards the simplest version. 

By contrast, the matrix for a hub model governs the behavior of the groups through the leader.  In the simplest version of hub models, there is no leader and nodes should appear in groups independently based on a multivariate Bernoulli distribution. But $A_{ij}=0$ means that if $v_i$ is the leader of a group, then $v_j$ will not appear in the group.  This implies a negative correlation between $v_i$ and $v_j$. Therefore, we do not directly penalize $A$ by $L_1$ penalization. We show in the following section that PCHM  penalizes $A$ towards the correct simplest version for hub models. 

\end{remark}

\section{Methodology}\label{S:Methodology}

\subsection{Penalized Likelihood}\label{S:likelihood}

The likelihood function of the hub model can be written as

\begin{equation}\label{E:GSM ProbOfGt}
	\mathbb{P}(G|A,\rho)=\prod_t \left \{ \sum_i \rho_i G_i^{(t)}\prod_j A_{ij}^{G_j^{(t)}}(1-A_{ij})^{(1-G_j^{(t)})} \right \}.
\end{equation}

$\mathbb{P}(G^{(t)}|A,\rho)$ belongs to the family of finite mixture models, where $\rho_i$ is the mixing distribution and $G_i^{(t)}\prod_j A_{ij}^{G_j^{(t)}}(1-A_{ij})^{(1-G_j^{(t)})}$ is the component distribution \citep{Hastie09}. 

Observe that the simplest version of \eqref{E:GSM ProbOfGt} is the one where the model has only one component. Without loss of generality, this means that $\rho_n=1$ and $\rho_i=0$, for $i=1,...,n-1$. Under this assumption $\mathbb{P}(G^{(t)}|A,\rho)$  becomes, 

\begin{equation}\label{E:Null}
\mathbb{P}(G^{(t)}|A,\rho) = \prod_{j=1}^{n} \tilde{A}_{j}^{G_j^{(t)}}(1-\tilde{A}_{j})^{(1-G_j^{(t)})}, \notag
\end{equation}

where $\tilde{A}_{j}=A_{nj}$. 

The simplest version of the hub model implies that all the nodes behave independently, which suggests that regularization on $\rho_i$ can penalize the model towards the simplest version. 

We propose the following penalized likelihood to reduce the number of components in \eqref{E:GSM ProbOfGt},

\begin{equation}\label{E:PCHM ProbOfGt}
	f(G|A,\rho,\eta)=\prod_t \left \{ \sum_i \rho_i^\eta G_i^{(t)}\prod_j A_{ij}^{G_j^{(t)}}(1-A_{ij})^{(1-G_j^{(t)})}  \right \} ,
\end{equation}
where $\eta$ is a tuning parameter greater than 1.  We refer to \eqref{E:PCHM ProbOfGt} as the \textit{penalized component hub model} (PCHM).

When maximizing PCHM \eqref{E:PCHM ProbOfGt}, we use the maximum likelihood estimator (MLE) of HM, $\hat{\rho}^{HM}$, as the starting point. When the true value of $\rho_i$ is 0 or very small, $\hat{\rho}^{HM}_i$ is also close to 0. Then the probability that a node is the leader of any group is also close to 0, thus the $\hat{\rho}_i$ estimated by PCHM will be shrunken towards 0 quickly due the power $\eta$ in \eqref{E:PCHM ProbOfGt}. We will carefully explain the motivation of PCHM in the next subsection.

It is worth highlighting the fact that the penalized component hub model \eqref{E:PCHM ProbOfGt} does not belong to the classical ``loss+penalty'' framework (i.e., it is not a linear combination of the log-likelihood function and a penalty term). The reason we do not use this type of criterion is due to the existence of the inequality constraints $ 0 \leq  \rho_i \leq 1$. The ``loss+penalty'' type of criterion usually leads to a non-convex optimization problem with inequality constraints, which is difficult to solve. By contrast, we will show that \eqref{E:PCHM ProbOfGt} can be solved by an algorithm very similar to EM. Therefore, the complexity of the algorithm is same as the algorithm for HM.

\subsection{Optimizing the Penalized Component Hub Model}\label{S: Optimizing}

We now derive the estimating equations to optimize the penalized likelihood function \eqref{E:PCHM ProbOfGt}. It is very straightforward to write the Lagrangian function which enforces the condition that $\sum_i \rho_i=1$ and $A$ is symmetric.
\begin{equation}\label{E:PCHM Objective Function}
	\Lambda(G|A,\rho,\eta)=\log f(G|A,\rho,\eta)-\lambda_o  [(\sum_i \rho_i)-1  ]-\sum_{i<j}\lambda_{ij}(A_{ij}-A_{ji}).\notag
\end{equation}

Since our immediate interest is on $\rho$, we begin by taking the derivative with respect to $\rho$.

\begin{align}
	\frac{\partial}{\partial \rho_x} \Lambda(G)&=\sum_t \frac{ \eta \rho_x^{\eta-1} G_x^{(t)} \prod_{j}{A_{xj}^{G_j^{(t)}} (1-A_{xj})^{1-G_j^{(t)}}}}{\sum_i \rho_i^\eta G_i^{(t)}\prod_j A_{ij}^{G_j^{(t)}}(1-A_{ij})^{(1-G_j^{(t)})}}-\lambda_o\notag\\
	&= \sum_t \frac{1}{\rho_x}\frac{ \eta \rho_x^{\eta} G_x^{(t)} \prod_{j}{A_{xj}^{G_j^{(t)}} (1-A_{xj})^{1-G_j^{(t)}}}}{\sum_i \rho_i^\eta G_i^{(t)}\prod_j A_{ij}^{G_j^{(t)}}(1-A_{ij})^{(1-G_j^{(t)})}}-\lambda_o.\notag
\end{align}

Applying stationary conditions, 

\begin{align} 
\rho_x&=\frac{1}{\lambda_o}\sum_t \frac{ \eta \rho_x^{\eta} G_x^{(t)} \prod_{j}{A_{xj}^{G_j^{(t)}} (1-A_{xj})^{1-G_j^{(t)}}}}{\sum_i \rho_i^\eta G_i^{(t)}\prod_j A_{ij}^{G_j^{(t)}}(1-A_{ij})^{(1-G_j^{(t)})}}.\notag
\end{align}

Before preceding we solve for $\lambda_o$. Noticing that $\sum_x \rho_x=1$, it is easy to obtain $\lambda_o=\eta T$. 

Thus, 
\begin{align}
\rho_x=\sum_t \frac{1}{T}\frac{\rho_x^{\eta} G_x^{(t)} \prod_{j}{A_{xj}^{G_j^{(t)}} (1-A_{xj})^{1-G_j^{(t)}}}}{\sum_i \rho_i^\eta G_i^{(t)}\prod_j A_{ij}^{G_j^{(t)}}(1-A_{ij})^{(1-G_j^{(t)})}}.\notag
\end{align}

With some minor abuse of notation, let
\begin{equation}\label{E:PCHM E-step}
f(S_x^{(t)}|G^{(t)})=\frac{\rho_x^{\eta} G_x^{(t)} \prod_{j}{A_{xj}^{G_j^{(t)}} (1-A_{xj})^{1-G_j^{(t)}}}}{\sum_i \rho_i^\eta G_i^{(t)}\prod_j A_{ij}^{G_j^{(t)}}(1-A_{ij})^{(1-G_j^{(t)})}}.
\end{equation}

This gives an estimating equation for $\rho_x$ of
\begin{equation}\label{E:PCHM rho_hat}
\hat{\rho}_x=\sum_t \frac{f(S_x^{(t)}|G^{(t)})}{T}.
\end{equation}
By applying similar techniques and those outlined in \citet{Zhao16}, it is straightforward to show that
\begin{equation}\label{E:PCHM A_hat}
	\hat{A}_{xy}=\frac{\sum_t G_y^{(t)} f(S_x^{(t)}|G^{(t)})+\sum_t G_x^{(t)}f(S_y^{(t)}|G^{(t)})}{\sum_t \big[f(S_x^{(t)}|G^{(t)})+f(S_y^{(t)}|G^{(t)})\big]}.
\end{equation}

The above estimating equations suggest an algorithm which is virtually identical to an EM algorithm.  That is, we iteratively perform an ``E-step'' by solving \eqref{E:PCHM E-step} then perform an ``M-step'' by solving \eqref{E:PCHM rho_hat} and \eqref{E:PCHM A_hat}. We refer to this as a \textit{pseudo-Expectation Maximization algorithm} because it is not maximizing the likelihood function, and \eqref{E:PCHM E-step} is not actually a posterior ``probability''; however, the performance is similar. 

Recall from Section \ref{S: GenMech} that there is an unobserved matrix, $S$, which indicates the true leader of observed groups.  Under the HM, the EM algorithm estimates $S$ as a $T \times n$ matrix during the E-step.  Each row of this posterior probability matrix sums to one.

The key feature of our approach to penalizing $\rho$ is that \eqref{E:PCHM E-step} also produces a $T \times n$ matrix where every row sums to one. Additionally, the EM algorithm shrinks $\rho_x$ by  penalizing the posterior probability $\mathbb{P}(S_x^{(t)}|G^{(t)})$, which will lead to some unusual behaviors. These behaviors will be demonstrated and discussed in Section \ref{S:Data}.  

Algorithm \ref{A:EM Algorithm} shows the details of the pseudo-EM algorithm. We always use the MLE $\hat{A}^{HM}$ and $\hat{\rho}^{HM}$ from HM as the starting point of PCHM. To obtain $\hat{A}^{HM}$ and $\hat{\rho}^{HM}$, we run the standard EM algorithm using a number of random starting points. We use 20 starting points in the simulation studies and 100 starting points in the data analysis. 

Additionally, note that the right hand side of \eqref{E:PCHM E-step} is always between 0 and 1. So the updated $\hat{\rho}_x$ satisfies the inequality constraint automatically. Therefore, the pseudo-EM algorithm does not require special considerations to satisfy the inequality constraints.

PCHM can produce a sparse estimator $\hat{\rho}$ for two reasons. Firstly, zero is an absorbing state for $\hat{\rho}$ in this algorithm.  It is easy to show that if at the $m^{th}$ iteration, $\hat{\rho}_x^{(m)}=0$, then by  \eqref{E:PCHM E-step} $f(S_x^{(t)}|G^{(t)})=0$ and by \eqref{E:PCHM rho_hat} $\hat{\rho}_x^{(m+1)}=0$.  Therefore, $\hat{\rho}_x=0$ is an absorbing state. Secondly, when $\hat{\rho}_x^{(m)}$ is small, $\mathbb{P}(S_x^{(t)}|G^{(t)})$ is also small and the power $\eta$ in \eqref{E:PCHM E-step} will make $\hat{\rho}_x$ quickly converge to 0. 

Due to numerical imprecision, Algorithm \ref{A:EM Algorithm} cannot guarantee that $\rho_i$ will reach absolute zero.  In practice, we set $\hat{\rho}_i=0$ if the value is less than $10^{-6}$. The estimates are in fact very insensitive to this threshold.

\begin{singlespace}
\begin{algorithm}[H]
 \textbf{Input}: $G$, $\eta$ \\
 \textbf{Output}: $\hat{A}, \hat{\rho}$
 
 \textbf{Initialize:}\\

$\>$ $loopBound=5000$\\

 $\>$	$A^{(0)}= \hat{A}^{HM}$ \\
$\>$  $\rho^{(0)}=\hat{\rho}^{HM}$ \\
$\>$ reps=starts
 
 	$counter=0$\\
 	
 	\While{	$\frac{\Delta \mathcal{L}(G|A^{(m+1)})}{\mathcal{L}(G|A^{(m)})}>10^{-6}$}{
 			\textbf{E-Step}\\
 			$\>$ Update $f(S_k^{(t)}=1|G^{(t)})$ by Equation \ref{E:PCHM E-step}\\
 			\textbf{M-Step}\\
 			$\>$ Update $A^{(m+1)}$ by Equation \ref{E:PCHM A_hat}\\
 			$\>$ Update $\rho^{(m+1)}$ by Equation \ref{E:PCHM rho_hat}\\
 			$\Delta \mathcal{L}(G|A^{(m+1)})=\mathcal{L}(G|A^{(m+1)})-\mathcal{L}(G|A^{(m)})$\\
 			$counter=counter+1$\\
 				\If{$counter>loopBound$}{
 					$\Delta \mathcal{L}(G|A^{(m+1)})=0$
 				}
 	}

  \caption{Pseudo-EM for PCHM}
  \label{A:EM Algorithm}
 \end{algorithm} 
 \end{singlespace}

\section{Selecting Tuning Parameters}\label{S:Selecting Tuning Parameters}
In this section, we select the tuning parameter $\eta$ which minimizes the Bayesian Information Criterion (BIC).

The BIC is frequently used to fit a model using maximization of a log-likelihood function \citep{Hastie09}.  The generic form of BIC is:

\begin{equation}\label{E:BIC}
	BIC = -2 \mathcal{L} + (\log T) d,
\end{equation}

\noindent where $\mathcal{L}$ is the log-likelihood, $T$ is the number of observations, and $d$ is the number of parameters. 

%The BIC is a criterion which attempts to balance the reduction in parameters achieved by penalizing the model against the decrease in likelihood which is caused by not using the maximum likelihood estimates.  That is, reducing parameters introduces a systematic error, \textit{bias}, into the estimates.  So long as the decrease in parameters is offset against the increase in bias, the BIC will go down.  Thus, the ``optimal'' model occurs at the minimum BIC.

Researchers often use the BIC to balance the goodness-of-fit against the complexity of a model. The second term of the criterion penalizes the number of parameters and thus can avoid the selection of the full models. 

In previous sections, we have calculated the number of parameters in the HM as 
	$d=[\frac{n(n-1)}{2}+(n-1)]$.
This is the number of unique symmetric elements in $A$, $\frac{n(n-1)}{2}$, according to the symmetry assumption, and the number of free parameters in $\rho$, $(n-1)$. 

However, as the number of non-zero elements in $\rho$ decreases from $n$, the structure of $A$ changes and $d$ will not follow this equation (see discussion in \ref{S:PCHM Motivation}).  Recalling that $n_o$ is the number of non-zero elements in $\rho$, the formula for the number of parameters becomes:

\begin{equation}\label{E:Parameters_PCHM}
	d=\underbrace{\frac{n_o(n_o-1)}{2}}_\text{X}+\underbrace{n_o(n-n_o)}_\text{Y}+\underbrace{(n_o-1)}_\text{$\rho$}.
\end{equation}

From \eqref{E:Parameters_PCHM}, when $n_o$ is close to $n$, the reduction in parameters is not very significant.  However, as the number of non-zero nodes decreases, the number of parameters begins to decline rapidly.  This suggests that while PCHM will be beneficial in simplifying models when a significant number of $\rho$'s are reduced to zero, there may be little benefit if the dataset contains many nodes which cannot be reduced to zero.  

This has implications for the types of datasets where PCHM can be employed.  In Section \ref{S:Review}, we pointed out that the assumption that $A_{ii}=1$ means that for groups where node $v_i$ is observed as a singleton, $v_i$ is the center.  Thus, $\rho_i$ must be greater than zero.  Therefore in datasets where singletons are common or the group size is small, the degree to which $n_o$ can be reduced is limited.

\section{Simulation Studies}\label{S:Simulation}

To better understand the performance of PCHM and demonstrate some aspects of the approach, we perform a series of simulation studies.  We begin with a simple toy example to introduce some basic behaviors.  Then we conduct simulations with varying link density, strength of relationships, and sparsity to demonstrate how these aspects effect model performance.  Finally, we show that PCHM is a robust technique even when the assumptions of the hub model are not strictly valid.  

\subsection{Toy Example} %a_fminconRho_A7

First, we consider the following simple situation.  We have a set of parameters represented in Table \ref{T:PenaltyExample} where there are only two members of the population exerting leadership.  We have a limited set of 20 observations shown in Table \ref{T:PenaltyObservations}. Notice that node $v_4$ is not a leader, but it is very popular.

\begin{table}[htb!]
\caption{Example of a Model With Sparse $\rho$'s (Rows of adjacency matrix with nonzero $\rho$'s are shown). }
\label{T:PenaltyExample}
\begin{center}
\begin{tabular}{|l | c | l l l l l l l|}
\hline
& & \multicolumn{7}{c|}{$j$} \\
\hline
$\rho$ & $i$ & 1 & 2 & 3 & 4 & 5 & 6 & 7\\
\hline
0.5 & 1 & 1.0000 & 0.7854 & 0.0000 & 0.9063 & 0.0000 & 0.0000 &  0.7452 \\
0.5 & 2 & 0.7854 & 1.0000 & 0.8324 & 0.8817 & 0.5885 & 0.8594 &  0.0000 \\
\hline

\end{tabular}
\end{center}
\end{table}

 \begin{table}[htb!]
 \caption{Frequency Table Low Observations Example}
 \label{T:PenaltyObservations}
  \begin{center}
  \begin{tabular}{| l l l l l l l | c | c |}
  \hline
 \multicolumn{7}{|c|}{$G$} &  &\\
   \hline
 1 & 2 & 3 & 4 & 5 & 6 & 7& Frequency & Elements\\
 \hline
1 & 0 & 0 & 0 & 0 & 0 & 0 & 1 &  1 \\
1 & 0 & 0 & 1 & 0 & 0 & 0 & 1 &  2 \\
1 & 1 & 0 & 0 & 0 & 0 & 1 & 1 &  3 \\
1 & 1 & 0 & 1 & 0 & 0 & 0 & 1 &  3 \\
0 & 1 & 1 & 1 & 1 & 0 & 0 & 2 &  4 \\
1 & 1 & 0 & 1 & 0 & 0 & 1 & 3 &  4 \\
1 & 1 & 0 & 1 & 0 & 1 & 0 & 1 &  4 \\
1 & 1 & 1 & 1 & 0 & 0 & 0 & 1 &  4 \\
1 & 1 & 0 & 1 & 1 & 1 & 0 & 1 &  5 \\
1 & 1 & 1 & 0 & 1 & 1 & 0 & 1 &  5 \\
1 & 1 & 1 & 1 & 0 & 1 & 0 & 5 &  5 \\
1 & 1 & 1 & 1 & 1 & 0 & 0 & 1 &  5 \\
1 & 1 & 1 & 1 & 1 & 1 & 0 & 1 &  6 \\
 \hline
 \multicolumn{7}{|c|}{Number of Times Observed}& 20 & \\
 \hline
  18 & 18 & 11 & 17 & 6 & 9 & 4 & & \\
 \hline
  \end{tabular}
  \end{center}
  \end{table}  

Table \ref{T:PenaltyRhos} shows the effect of applying the PCHM algorithm with increasing $\eta$ until the number of non-zero elements of $\rho$ has been reduced to $n_o=2$.  Notice that when we apply HM (i.e., $\eta=1$), the model estimates that nodes $v_1$ though $v_4$  all have non-zero $\rho$.  This is not surprising when we consider that each of these nodes appear in over half of the observations (see Table \ref{T:PenaltyObservations}).

As $\eta$ increases, $v_4$ is the first node to shrink to zero.  This reduction is achieved with a relatively low value of $\eta=1.1$.  When $\eta$ is increased to 1.7, the PCHM identifies the true sparsity of the original model and the Bayesian Information Criterion achieves its minimum.

\begin{table}[H]
\caption{Bayesian Information Criterion as $\eta$ Increases}
\label{T:PenaltyRhos}
\begin{center}
\resizebox{6in}{!}{
\begin{tabular}{|c|l l l l l l l | c | c | c |}
\hline
&\multicolumn{7}{c|}{$\rho_i$} & & &\\
\hline
$\eta$ & 1 & 2 & 3 & 4 & 5 & 6 & 7 & LL & BIC& $n_o$\\
\hline
1.0 & 0.3500 & 0.4507 & 0.0799 & 0.1194 & 0.0000 & 0.0000 & 0.0000 & -54.6946 & 172.2996 &  4 \\
1.1 & 0.3453 & 0.5597 & 0.0949 & 0.0000 & 0.0000 & 0.0000 & 0.0000 & -54.9719 & 160.8712 &  3 \\
1.2 & 0.3451 & 0.5612 & 0.0938 & 0.0000 & 0.0000 & 0.0000 & 0.0000 & -54.9730 & 160.8734 &  3 \\
1.3 & 0.3447 & 0.5630 & 0.0922 & 0.0000 & 0.0000 & 0.0000 & 0.0000 & -54.9756 & 160.8787 &  3 \\
1.4 & 0.3444 & 0.5655 & 0.0902 & 0.0000 & 0.0000 & 0.0000 & 0.0000 & -54.9813 & 160.8900 &  3 \\
1.5 & 0.3439 & 0.5689 & 0.0872 & 0.0000 & 0.0000 & 0.0000 & 0.0000 & -54.9935 & 160.9145 &  3 \\
1.6 & 0.3433 & 0.5744 & 0.0823 & 0.0000 & 0.0000 & 0.0000 & 0.0000 & -55.0242 & 160.9758 &  3 \\
\hline
1.7 & 0.3386 & 0.6614 & 0.0000 & 0.0000 & 0.0000 & 0.0000 & 0.0000 & -57.8882 & 151.7253 &  2 \\
\hline
1.8 & 0.3379 & 0.6621 & 0.0000 & 0.0000 & 0.0000 & 0.0000 & 0.0000 & -57.8896 & 151.7279 &  2 \\
1.9 & 0.3370 & 0.6630 & 0.0000 & 0.0000 & 0.0000 & 0.0000 & 0.0000 & -57.8913 & 151.7313 &  2 \\
2.0 & 0.3361 & 0.6639 & 0.0000 & 0.0000 & 0.0000 & 0.0000 & 0.0000 & -57.8933 & 151.7355 &  2 \\
\hline

\end{tabular}
}
\end{center}
\end{table}

It is worth noticing that the lowest value that $n_o$ can take on is two.  To see why this is true, consider Table \ref{T:PenaltyObservations}.  As mentioned earlier, if a node appears as a singleton, there must be a non-zero $\rho$ associated with it.  From this, we might expect that the minimum value of $n_o$ would be one because there is only one singleton in the dataset.  However, notice that there is one group (the fifth row of Table \ref{T:PenaltyObservations}) which does not have $v_1$ as a member.  Thus, $v_1$ could not have created this observation and there must be at least one more node with a non-zero $\rho$.

\begin{table}[htb!]
\caption{Rows of Estimated Adjacency Matrix with Nonzero $\rho$'s for Optimal $\eta$}
\label{T:PenaltyA_1.7}
\begin{center}
\begin{tabular}{|l | c | l l l l l l l|}
\hline
& & \multicolumn{7}{c|}{$j$} \\
\hline
$\rho$ & $i$ & 1 & 2 & 3 & 4 & 5 & 6 & 7\\
\hline
0.339 & 1 & 1.000 & 0.800 & 0.000 & 0.705 & 0.000 & 0.000 &  0.591 \\
0.661 & 2 & 0.800 & 1.000 & 0.832 & 0.924 & 0.454 & 0.680 &  0.000 \\
\hline

\end{tabular}
\end{center}
\end{table}

One might observe that the estimates in Table \ref{T:PenaltyA_1.7} are not very accurate when compared to the true adjacency matrix (Table \ref{T:PenaltyExample}); however, with only 20 observations, it is not surprising that the estimates are not very accurate.

\begin{remark}
In Table \ref{T:PenaltyRhos}, we notice that when the number of parameters decreases, there is a sudden drop in BIC.  However, observe that, in general, the lowest value of $\eta$ that produces a certain number of parameters also seems to be the $\eta$ associated with the lowest BIC.  That is, BIC tends to increase as $\eta$ increases when the number of parameters remains the same.  This is because the estimation bias is increasing with no decrease in the number of parameters.  This is a natural result of \eqref{E:BIC}.
\end{remark}

\subsection{Estimating Parameters for Networks $(n=50)$ with link density 0.5}\label{S:large}

In this section, we randomly generate sparse networks of 50 nodes.  In each network $n_o=8$.  For each element in the networks we apply the following distribution:

\[ A_{ij} = \left\{ 
   \begin{array}{l l}
     Beta(\alpha,\beta) & \quad \textnormal{with probability $p$,}\\
     0 & \quad \textnormal{with probability $(1-p)$,}
   \end{array} \right.\]
where $p$ is the link density. For the initial simulation, only half of the pairs will have a relationship (i.e., $p=0.5$) and  the average relationship will be 0.25 (i.e., $\alpha=1$ and $\beta=3$). We fix $\rho$ to be $(1/n_o,...,1/n_o,0,...,0)$ for all the simulations. 

\subsubsection{Single Simulated Dataset}

\begin{figure}[H]
\centering
	\includegraphics[width=0.8\textwidth]{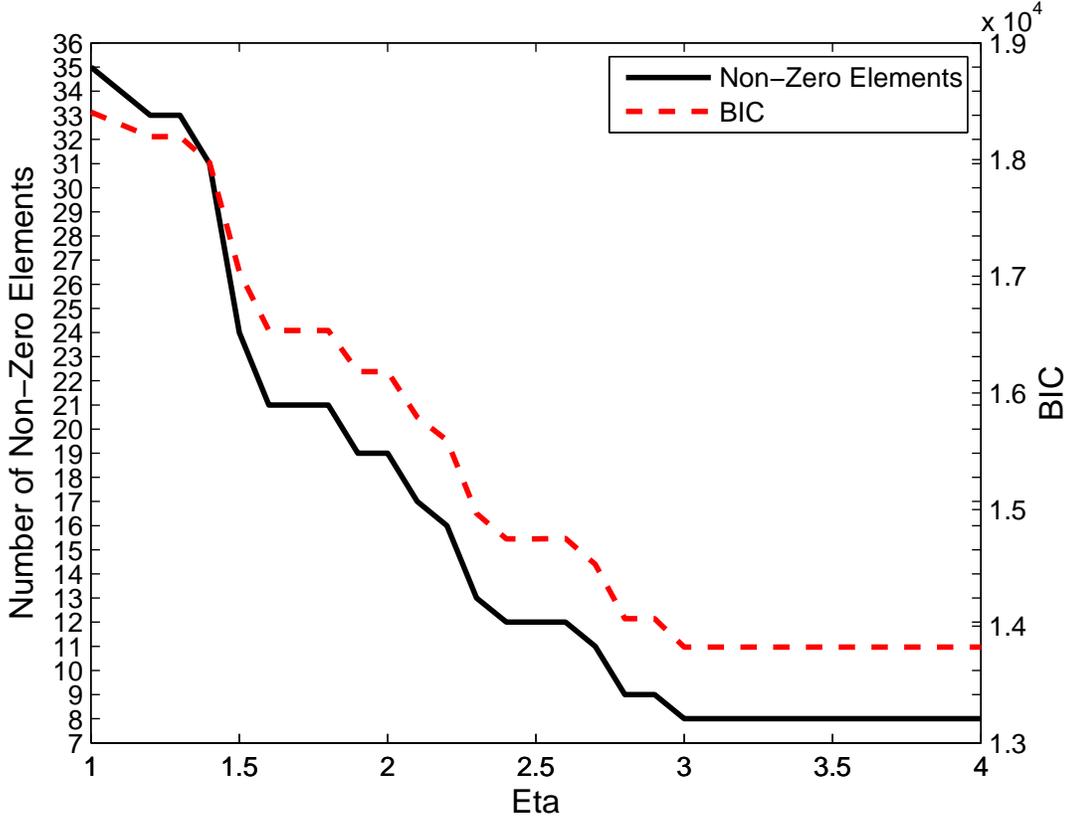}
	\caption{For a population of 50 nodes where only 8 members initiate groups, increasing the penalty, $\eta$, reduces the number of parameters and achieves the true number of non-zero elements of $\rho$ with the minimum BIC.}
	\label{F:Sim_50_8}
\end{figure}

For our first simulation, we create a single set of parameters and generate $T=500$ observations.  The results of this simulation are shown in Figure \ref{F:Sim_50_8} where we see that the number of non-zero elements in $\rho$ decrease until the minimum BIC is achieved at $\eta=3.0$.

In this graphical representation, we plot the BIC along with $n_o$.  It may appear to the naked eye that the BIC is not rising for $\eta>3.0$; however, it is increasing very slightly.

\subsubsection{General Simulation Results}\label{S:general}

%\citet{Zhao15} 
We use the mean absolute error (MAE) as a measure of the overall accuracy of an estimate of hub model parameters.
\begin{align}
	MAE(A)&=\frac{1}{\binom{n}{2}} \sum_{i<j} |\hat{A}_{ij}-A_{ij}|,\notag\\
	MAE(\rho)&=\frac{1}{n} \sum_{i} |\hat{\rho}_{i}-\rho_{i}|.\notag
\end{align}

For both HM and PCHM, we set $\hat{\rho}_i=0$ if the value is less than $10^{-6}$. In addition, when both $\hat{\rho}_i=0$ and $\hat{\rho}_j=0$, we set $\hat{A}_{ij}=0$, when we calculate the MAE. 

We generate 200 sets of parameters $(A,\rho)$ using the same technique described above. We generate a dataset with $T$ groups based on each $(A,\rho)$. We then fit HM and PCHM on each dataset and the result is summarized in Tables \ref{T:MAEs} and \ref{T:n0s}. Each average and standard deviation are calculated over the 200 datasets. To evaluate the performance of the methods under various sample sizes, we try 7 different values of $T$, that is $T=$100, 200, 500, 1000, 2000, 5000 and 10000. The tuning parameter $\eta$ is chosen based on a grid value search from 1 to 15 with increment 0.5.

%\cite{Zhao15} 
\citet{Zhao16} observed that as the number of observations increases, the MAE of the estimated $A$ improves.  However, for situations where there is a large number of nodes, the improvement rate can be very slow.  Since PCHM does not seem to introduce significant bias into the individual elements of $A$, we now show that for sparse latent network structure, the PCHM estimates are much more accurate than HM according to MAE. 

First, notice in Table \ref{T:MAEs} that as the number of observations increases, the error in $A$ and $\rho$ under PCHM is quickly shrunken towards zero while the error in $A$ under HM does not significantly improve despite a dramatic increase in the number of observations.  Also note that the MAE of the PCHM is always lower than that of HM.

\begin{table}[H]
\caption{Comparison of Estimation Error from Hub Model and Penalized Component Hub Model ($n_o=8$, link density=0.5)}
\label{T:MAEs}
\begin{center}
\resizebox{6in}{!}{
\begin{tabular}{|c | c | c || c | c |}
\hline

& \multicolumn{2}{c||}{MAE($A$)} & \multicolumn{2}{c|}{MAE($\rho$)}\\
\hline
& HM & PCHM& HM & PCHM \\
\hline
Obs & Avg (StDev) & Avg(StDev) & Avg(StDev) & Avg(StDev)\\
\hline
100 & 0.1118 (0.0129) & 0.0748 (0.0167) & 0.0251 (0.0045) & 0.0182 (0.0049)
 \\
200 & 0.0991 (0.0147) & 0.0345 (0.0149) & 0.0159 (0.0044) & 0.0078 (0.0038)
\\
500 & 0.0845 (0.0110) & 0.0066 (0.0042) & 0.0055 (0.0018) & 0.0021 (0.0011)

  \\
1000 & 0.0828 (0.0100) & 0.0041 (0.0018) & 0.0029 (0.0006) & 0.0014 (0.0005)

  \\
2000 & 0.0819 (0.0098) & 0.0028 (0.0005) & 0.0018 (0.0003) & 0.0010 (0.0003)

  \\
5000 & 0.0798 (0.0094) & 0.0019 (0.0011) & 0.0009 (0.0002) & 0.0006 (0.0002)
    
 \\
10000 & 0.0776 (0.0098) & 0.0013 (0.0005) & 0.0006 (0.0001) & 0.0004 (0.0001)
  
\\

\hline

\end{tabular}
}
\end{center}
\end{table}

Next, consider Table \ref{T:n0s} where we compare the number of non-zero elements in each estimated $\rho$ as well as the corresponding number of parameters $d$.  Not only does the HM not reduce the number of parameters in the estimates, but $n_o$ actually increases as the number of observations increases.  On the other hand, PCHM identifies a much sparser model and detects the true number of nodes influencing the group behavior with very few observations.

\begin{table}[H]
\caption{Comparison of Model Selection from Hub Model and Penalized Component Hub Model ($n_o=8$, link density=0.5)}
\label{T:n0s}
\begin{center}
\resizebox{6in}{!}{
\begin{tabular}{|c | c | c ||c|c||c|}
\hline

& \multicolumn{2}{c||}{Estimated $n_o$} & \multicolumn{2}{c||}{Estimated $d$} & $\eta$\\
\hline
& HM & PCHM & HM & PCHM & PCHM\\
\hline
Obs & Avg (StDev) & Avg(StDev) & Avg (StDev) & Avg(StDev) & Avg(StDev)\\
\hline
100  & 31.2050 (2.1131) & 18.8950 (2.6777) & 503.7000 (67.3348) & 190.5250 (51.2374) & 10.4825 (2.7125)

 \\
200  &  32.7850 (2.4431) & 12.6200 (2.7465) & 555.7900 (82.1888) & 88.6950 (37.9984) & 9.6400 (2.5848)

\\
500  &  33.0400 (2.1544) & 8.2300 (0.7415) & 563.6500 (72.0767) & 37.2550 (7.6965) & 7.6050 (2.7311)

\\
1000 & 33.8850 (2.0378) & 8.0500 (0.3287) & 592.1050 (69.9353) & 35.4800 (3.3279) & 5.2375 (2.2822)

 \\
2000  & 34.2700 (1.8987) & 8.0200 (0.1404) & 605.1450 (66.1839) & 35.1800 (1.2632) & 3.8925 (1.4604)

 \\
5000  &  34.4100 (2.2555) & 8.0450 (0.2078) & 610.7600 (77.8296) & 35.4050 (1.8704) & 3.0750 (0.4216)

\\
10000 &  34.3450 (1.8878) & 8.0200 (0.1404) & 607.7350 (65.5346) & 35.1800 (1.2632) & 2.8475 (0.3294)

 \\

\hline

\end{tabular}
}
\end{center}
\end{table}

Finally, the last column of Table \ref{T:n0s} shows the average value of $\eta$ which achieves the minimal BIC.  Notice that as the number of observations increases, the optimal penalty decreases.  Since the true parameters are sparse, as the number of observations gets larger, the starting point gets closer to the true parameters and less penalty is needed. In the supplemental materials, we will perform more simulation studies under various parameter settings.

\subsection{Testing robustness of PCHM} 

The hub model makes two key assumptions: 

\begin{enumerate}
\item The groups are generated independently from the previous groups.
\item There is only one leader in each group.
\end{enumerate}

Both assumptions can be violated in some practical examples. If groups are ordered in time, there may exist dependency among groups. Also there may exist multiple leaders in some groups. 

In this section, we test the robustness of PCHM, especially its ability to identify the leader set if the data is not generated exactly by the hub model. 

\subsubsection{Dependency Among Groups}

We first consider the effect of dependency between some groups. We employ a time varying hub model proposed by 
one of the authors. This model assumes that the dataset can be partitioned into multiple time segments. The groups in different segments are independent while they have Markov dependence of order 1 within the same time segments. 

Specifically, a group $G^{(t)}$ at time $t$ is a start of a new segment with probability $q$, and is within the same segment as the previous group with probability $1-q$. In the former case, the group is generated by the hub model. For the purpose of this article, we continue to assume the sparsity of $\rho$ in the hub model, i.e., $\rho=(1/n_o,...,1/n_o,0,...,0)$. However, given the group leader, $A_{ij}=\frac{\exp\{ \theta_{ij}\}}{1+\exp\{ \theta_{ij}\}}$. 

When an observation is not the start of a new segment, $G^{(t)}$ is considered as a transformation of $G^{(t-1)}$. That is, the leader in $G^{(t)}$ is the same as in $G^{(t-1)}$. If $G^{(t-1)}_j=1$, the leader will continue to include $v_j$ in the group with probability 
$
 B_{ij}= \frac{\exp\{ \theta_{ij}+\gamma_b \}}{1+\exp\{ \theta_{ij}+\gamma_b \}}
$. If $G^{(t-1)}_j=0$, the leader will choose to add $v_j$ to the group with probability
$
 C_{ij}= \frac{\exp\{ \theta_{ij}+\gamma_c \}}{1+\exp\{ \theta_{ij}+\gamma_c \}}
$. When $\gamma_b>0$ and $\gamma_c<0$, this setup suggests some level of the stability among groups within the same segment. That is, a group member has a higher probability of being kept in the group and an outsider has a lower probability of being selected into the group. 

For our simulation, we generate $A$ from the model defined in Section \ref{S:large} and obtain $\Theta=[\theta_{ij}]$ correspondingly. We set $q=0.2$, $\gamma_b=1$ and $\gamma_c=-1$. We generate $G$ from the same procedure described in Section \ref{S:large}.

The results are given in Tables \ref{T:MAEsTime} and \ref{T:n0sTime}. The results are naturally less impressive than the results in Tables \ref{T:MAEs} and \ref{T:n0s} because the model was correctly specified in the earlier simulation. However, PCHM still outperforms HM in both parameter estimation and leader set identification. Specifically, PCHM still results in a much sparser model and eventually approaches the correct $n_o$. 

\begin{table}[H]
\caption{Comparison of Estimation Error from Hub Model and Penalized Component Hub Model for Time Varying Groups}
\label{T:MAEsTime}
\begin{center}
\resizebox{6in}{!}{
\begin{tabular}{|c | c | c || c | c |} 
\hline

& \multicolumn{2}{c||}{MAE($A$)} & \multicolumn{2}{c|}{MAE($\rho$)}\\
\hline
& HM & PCHM& HM & PCHM \\
\hline
Obs & Avg (StDev) & Avg(StDev) & Avg(StDev) & Avg(StDev)\\
\hline
100 & 0.2118 (0.0159) & 0.1773 (0.0205) & 0.0317 (0.0037) & 0.0300 (0.0045)
 \\
200 & 0.2061 (0.0152) & 0.1417 (0.0235) & 0.0279 (0.0043) & 0.0241 (0.0047)

\\
500 & 0.1823 (0.0147) & 0.0749 (0.0261) & 0.0166 (0.0041) & 0.0137 (0.0042)

  \\
1000 & 0.1725 (0.0123) & 0.0430 (0.0226) & 0.0089 (0.0026) & 0.0068 (0.0032)

  \\
2000 & 0.1703 (0.0114) & 0.0260 (0.0194) & 0.0053 (0.0019) & 0.0038 (0.0019)

  \\
5000 & 0.1710 (0.0113) & 0.0185 (0.0187) & 0.0031 (0.0017) & 0.0024 (0.0016)

 \\
10000 & 0.1736 (0.0116) & 0.0149 (0.0137) & 0.0020 (0.0013) & 0.0016 (0.0011)

\\

\hline

\end{tabular}
}
\end{center}
\end{table}

\begin{table}[H]
\caption{Comparison of Model Selection from Hub Model and Penalized Component Hub Model for Time Varying Groups}
\label{T:n0sTime}
\begin{center}
\resizebox{6in}{!}{
\begin{tabular}{|c | c | c ||c|c||c|}
\hline

& \multicolumn{2}{c||}{Estimated $n_o$} & \multicolumn{2}{c||}{Estimated $d$} & $\eta$\\
\hline
& HM & PCHM & HM & PCHM & PCHM\\
\hline
Obs & Avg (StDev) & Avg(StDev) & Avg (StDev) & Avg(StDev) & Avg(StDev)\\
\hline
100  & 32.3800 (2.4072) & 22.6050 (3.0040) & 542.3050 (79.3144) & 270.2850 (68.7773) & 10.9425 (2.7842)
 \\
200  & 36.3950 (2.1801) & 18.9650 (3.5022) & 681.8600 (80.7547) & 194.4200 (69.4177) & 10.7650 (2.9012)

\\
500  & 39.2300 (2.0043) & 13.0750 (3.2111) & 790.1100 (79.4202) & 96.1450 (47.4858) & 7.9700 (2.3594)

\\
1000 & 40.1150 (1.8760) & 10.9300 (2.5013) & 825.4150 (76.0390) & 67.3100 (31.4978) & 6.3275 (1.3086)

 \\
2000  & 41.4750 (1.8375) & 9.3950 (2.0762) & 881.5050 (77.3807) & 49.9750 (25.3889) & 5.6825 (1.0776)

 \\
5000  & 43.4350 (1.6672) & 8.7550 (2.0582) & 965.4000 (72.8805) & 43.8100 (26.7504) & 4.6675 (1.0553)

\\
10000 & 43.9661 (1.5740) & 8.4576 (1.6623) & 988.7203 (69.6916) & 40.3644 (20.7384) & 4.1017 (0.5569)

 \\

\hline

\end{tabular}
}
\end{center}
\end{table}

\subsubsection{Multiple Leaders}

In this final simulation, we consider groups with multiple leaders. We test our method on a simple model with two leaders for each group. Two leaders are sampled from $\rho=(1/n_o,...,1/n_o,0,...,0)$ without replacement. With $v_i$ and $v_j$ as the group leaders, $\mathbb{P}(G_k^{(t)}=1|S_i^{(t)}=1,S_j^{(t)}=1)= \frac{\exp\{ h_{ik}+h_{jk}+c \}}{1+\exp\{h_{ik}+h_{jk}+c \}}$, where $H=[h_{ik}]$ is an $n_o$ by $n$ matrix with $h_{ik} \geq 0$. 

In this model, the appearance of node $v_k$ in a group is influenced by both leaders through $h_{ik}$ and $h_{jk}$. We assume that all components in $H$ are non-negative to make both leaders have positive effect on attraction of group members. 

We set $n_o=8$, generate each $h_{ik}$ from $U(0,1)$ independently, and choose $c=-3.5$ to adjust the average group size to be comparable with the setup in Section \ref{S:large}. 

The results are given in Table \ref{T:n0sMulti}. Since there is no reasonable comparison of $\hat{A}$ and $H$, we focus on leader set identification. Because the data generating procedure deviates from the hub model more than the time varying example, it is more difficult to detect the leader set using PCHM. However, PCHM still produces a sparser model and eventually identifies a leader set close to the truth. 

\begin{table}[H]
\caption{Comparison of Model Selection from Hub Model and Penalized Component Hub Model for Time Varying Groups}
\label{T:n0sMulti}
\begin{center}
\resizebox{6in}{!}{
\begin{tabular}{|c | c | c ||c|c||c|}
\hline

& \multicolumn{2}{c||}{Estimated $n_o$} & \multicolumn{2}{c||}{Estimated $d$} & $\eta$\\
\hline
& HM & PCHM & HM & PCHM & PCHM\\
\hline
Obs & Avg (StDev) & Avg(StDev) & Avg (StDev) & Avg(StDev) & Avg(StDev)\\
\hline
100  & 36.0650 (1.9129) & 26.1900 (2.5821) & 669.1950 (70.0464) & 358.3700 (69.0955) & 10.4225 (2.9664)

 \\
200  & 40.1050 (1.7519) & 25.1350 (3.2604) & 824.7850 (71.0863) & 332.7400 (83.5993) & 9.9275 (3.0832)

\\
500  & 43.0550 (1.6540) & 22.7600 (4.7163) & 948.7550 (71.8719) & 280.4550 (110.2504) & 8.7700 (3.3989)

\\
1000 & 44.0350 (1.5315) & 21.9050 (5.4979) & 991.7250 (68.1297) & 264.9050 (128.4688) & 7.8750 (3.7787)

 \\
2000  & 44.4950 (1.5204) & 20.0850 (7.6255) & 1012.3000 (68.0081) & 239.6750 (147.8264) & 6.2550 (3.4210)

 \\
5000  & 44.3316 (1.5785) & 7.9037 (4.5745) & 1005.0481 (70.8127) & 44.5936 (89.5893) & 3.6765 (0.5116)

\\
10000 & 43.8487 (1.7106) & 7.0672 (0.2515) & 983.7311 (75.4491) & 27.5378 (2.0118) & 3.2563 (0.2673)

 \\

\hline

\end{tabular}
}
\end{center}
\end{table}

Notice in Table \ref{T:n0sMulti} that PCHM seems to eventually estimate a leader set which is slightly smaller than the true number of leaders.  This is a consequence of the HM assumption that every group has only one leader.  PCHM is penalizing the data towards a single leader and it is likely that the influence of some leaders will be marginalized.  For example, if there are two leaders who frequently appear together, the hub model will treat one of them as the leader and the other as a follower.

\section{Data Analysis}\label{S:Data}

In this section, we analysis a dataset from the animal behavior literature and records the interaction of Hector's dolphins \citep{Bejder98}. In this dataset, there are a moderate number of observations compared to the number of parameters. Here there are $n=18$ individuals which form $T=40$ groups. In this case, $T>n$ but $T<\big[\frac{n(n-1)}{2}+(n-1)\big]$. 

An important point about this dataset is that there are 5 nodes which appear as singletons.  This means that this is the lower bound of $n_o$.

This dataset consists of observations of Hector's dolphins (\textit{Cephalorhynchus hectori}) in the inshore waters of the South Island of New Zealand taken over 1996-1997.  The full population is estimated to contain 50-70 individuals.

Hector's dolphins are most often observed in groups of two to eight individuals.  These groups often fuse together and split up over periods of several hours.  The researchers considered individuals associated if they were members of the same group or cluster of groups.  Groups of dolphins were considered part of the same cluster of groups if groups merged in the time span when observations were being taken.

Observations were recorded by photographs.  Photography is a noninvasive tool that is frequently used to study the social structure of cetaceans and other social animals.  Groups are defined entirely based on photographic records.  That is, individuals seen but not photographed are not included in the observed group \citep{Bejder98}.

\begin{figure}[H]
\centering
	\begin{subfigure}[b]{0.45\textwidth}
	\includegraphics[width=\textwidth]{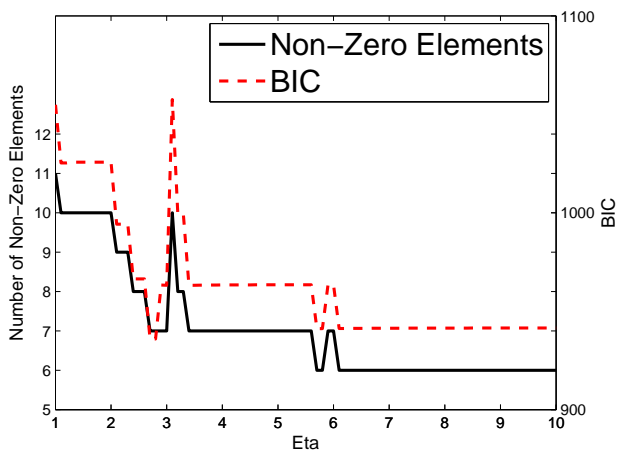}
		\caption{BICs and number of non-zero elements}
		\label{F:Dolphins}
	\end{subfigure}
	~
	\begin{subfigure}[b]{0.45\textwidth}
	\includegraphics[width=\textwidth]{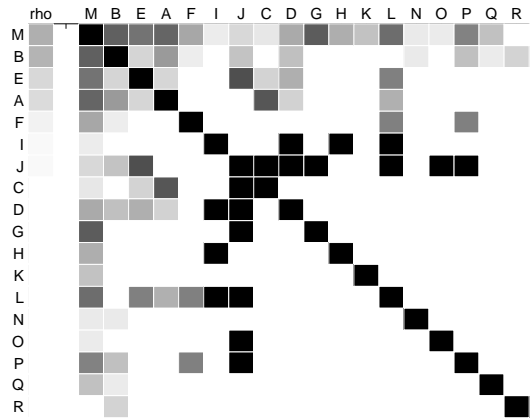}
		\caption{Grayscale plot}
		\label{F:Dolphins_Grey}
	\end{subfigure}
	\caption{In the Hector's dolphin dataset, subsets of 18 animals are observed in groups at 40 different independent times.}
\end{figure}

As with the simulations, we select $\eta$ by BIC. Here the optimal BIC is achieved at $\eta=2.7$, and there are 7 individuals in the population with non-zero $\hat{\rho}_i$. 

In Figure \ref{F:Dolphins_Grey}, we visualize the estimated adjacency matrix by a grayscale plot where the strength of a relationship is represented by the cell's color.  Nodes with weak relationships have light cells while nodes with strong relationships have dark cells.  Cells representing relationships of intermediate strength are shaded along the grayscale. In addition, the nodes are ranked in descending order of their $\hat{\rho}_i$.

Recall from above that there are five members of the population which appear as singletons.  These nodes are represented in Figure \ref{F:Dolphins_Grey} by the letters M, B, E, A, and F.  That is, the singletons account for all but the two nodes with the lowest $\hat{\rho}$.  In fact, nodes $v_I$ and $v_J$ have an estimated $\rho$ of $1/T$.

In Figure \ref{F:Dolphins}, we can see an additional effect of the penalization technique used in PCHM which is different from traditional methods of regularization. That is, the number of non-zero elements does not decrease monotonically. 

This behavior is related to the fact that the penalty is really affecting the E-step of the algorithm and the estimation of the posterior probability matrix.  The penalty induces sparsity in $\hat{S}$ and drives it towards a binary matrix.  Thus it is possible for $\hat{S}^{(\eta_1)}$ to be closer to a binary matrix than $\hat{S}^{(\eta_2)}$ but $n_o^{(\eta_1)}$ is greater than $n_o^{(\eta_2)}$.

While this behavior is not as desirable as monotonically declining elements, it does not appear that this negatively impacts the performance of the PCHM.

\begin{figure}[H]
\centering
	\begin{subfigure}[b]{0.45\textwidth}
	\includegraphics[width=\textwidth]{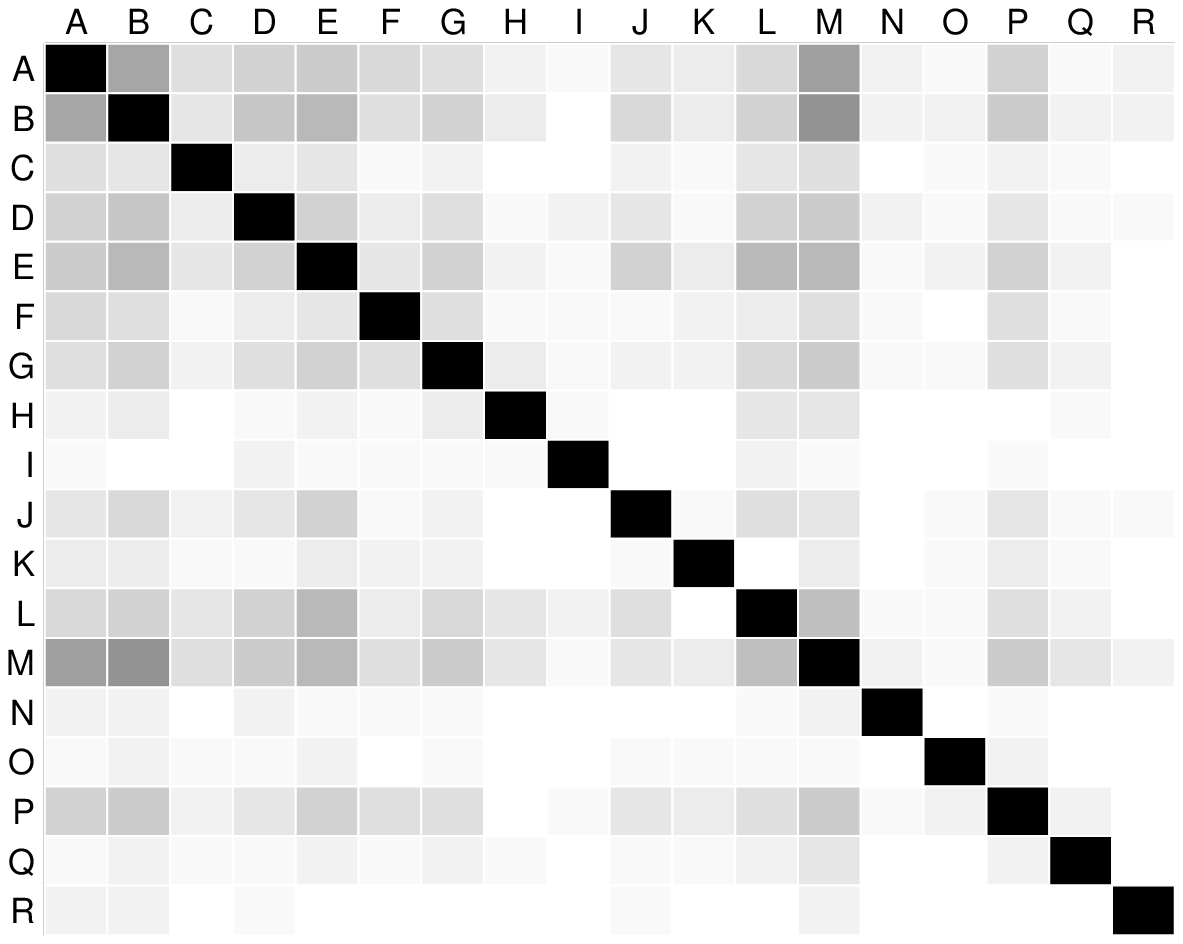}
		\caption{Co-occurrence Matrix}
		\label{F:Dolphins_O}
	\end{subfigure}
	~
	\begin{subfigure}[b]{0.45\textwidth}
	\includegraphics[width=\textwidth]{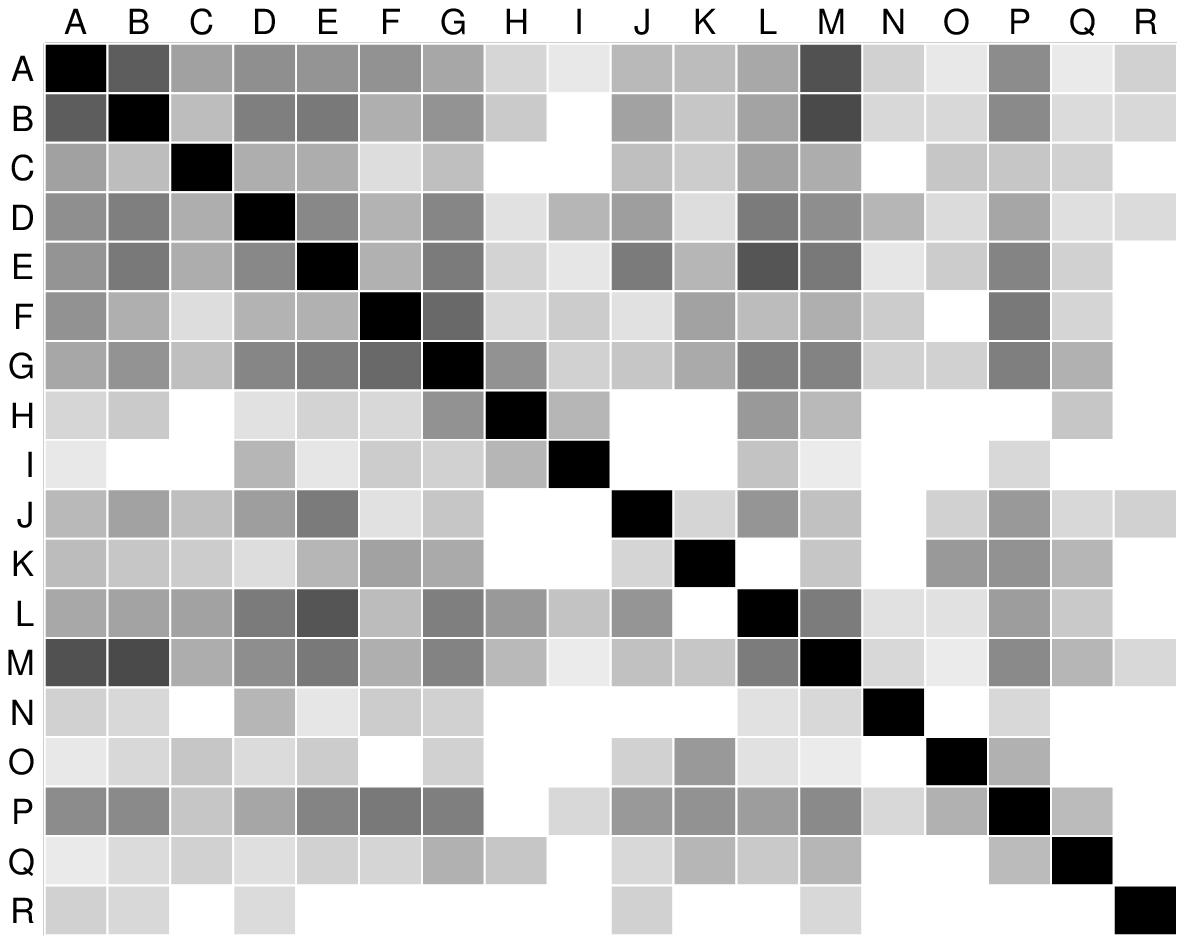}
		\caption{Half Weight Index}
		\label{F:Dolphins_H}
	\end{subfigure}
	\caption{Traditional measures of social networks applied to the Dolphin dataset}
\end{figure}

Figure 2 and 3 clearly demonstrate that PCHM provides a much simpler characterization of the population behavior than traditional measures.

In supplemental materials, we will analyze a dataset from the recommender system literature which provides user preferences \citep{Goldberg01}.

\section{Conclusion and Discussion}

In this article, we developed the penalized component hub model (PCHM) to improve the performance of the hub model (HM) proposed by the authors when applied to relatively small datasets. In previous work, it has been shown that datasets with relatively low numbers of observations tend to produce unstable estimates of the adjacency matrix. PCHM reduces this instability without introducing significant bias.  An additional benefit of PCHM is that even when the number of parameters is not larger than the sample size, analysts can benefit from reducing the complexity of the model. 

PCHM has two significant differences from the regularization penalization methods traditionally used for graph estimation. Firstly, PCHM penalizes $\rho_i$, the probability of nodes being centers, instead of $A_{ij}$ in the adjacency matrix. This method is conceptually appropriate  since it penalizes the model towards the simplest case - the case of independence.  Secondly, instead of using the ``loss+penalty'' paradigm, PCHM adds a power of $\eta$ on $\rho_i$ which may have more general applications to finite mixture models when the number of components is unknown. An advantage of this approach is that PCHM can be solved by the pseudo-EM algorithm very efficiently.

By applying the PCHM to Hector's Dolphins and Jester datasets, we demonstrate that PCHM obtains a sparse $\hat{\rho}$ and thereby reduces the model complexity when the sample size is moderate. In addition, PCHM only introduces mild estimation bias and thus the result is close to HM when the sample size is large.

While we have focused on social network analysis in this paper, this is primarily motivated by the intuitive link between social behavior and hub models.  In principle, hub models have application in any area which uses co-occurrence as a proxy for relationship strength.  Thus, market basket analysis, recommender systems, and biological networks all have potential for benefiting from hub models and PCHM.

Further, to the best of our knowledge, this approach to penalizing component selection in finite mixture models is itself a technical innovation. In Section \ref{S:likelihood}, we showed that regularization on $\rho_i$ rather than on $A_{ij}$ penalizes the model towards the simplest version, that is, every node behaves independently. However, there exists a small gap: since we assume $A_{nn}=1$, node $v_n$ must appear in every group to achieve the simplest version.  In future work, we consider relaxing this assumption and allow $A_{ii}$ to be an arbitrary number between 0 and 1. We will further investigate the theoretical properties of PCHM, for example, the behavior of the estimator as the size of the network grows. To fully understand the behavior of the pseudo-EM is also an intriguing topic. 

As mentioned in Section \ref{S:Review}, hub models and PCHM make two key assumptions: 1) independence among groups  2) one leader for each group. Both assumptions can be generalized for practical applications. In future work, we will generalize the idea of the hub model and PCHM to allow for flexible number of leaders and dependency structure among groups.

\appendix
\section*{Appendix A: Supplemental data}

Supplementary data associated with this article can be found in the online version at http://to-be-determined.

\end{document}